\documentclass[12pt,reqno,a4paper]{amsart}
\usepackage{amssymb,latexsym}
\usepackage{amsthm, amsmath, amsfonts}
\usepackage{hyperref}
\def\be{\begin{equation}}
\def\ee{\end{equation}}
\def\bea{\begin{eqnarray}}
\def\eea{\end{eqnarray}}
\def\bz{\bar z}
\def\bs{\bar s}
\def\p{\partial}
\def\bp{\bar\partial}

\def\Hilb{{\rm Hilb}}
\def\bs{\bar s}
\def \CO{\mathcal O}
\def\bR{\bar R}
\def\bcal_k{\mathcal B_k}
\newcommand{\szego}{Szeg\"o\ }
\newcommand{\kcalomega}{\mathcal{K}_{[\omega_0]}}

\begin{document}
%\hfill RUNHETC-2008-15\\

\title[Random normal matrices, Bergman kernel and projective embeddings
]{Random normal matrices, Bergman kernel
\vspace{.2cm}
and projective embeddings}
\author[Semyon Klevtsov]{Semyon Klevtsov}
\maketitle
\begin{center}
{\it\small Mathematisches Institut and Institut f\"ur Theoretische Physik,\\ Universit\"at zu K\"oln, Weyertal 86-90, 50931 K\"oln, Germany}
\end{center}
\vspace{.5cm}

\begin{abstract}
We investigate the analogy between the large $N$ expansion in normal matrix models and the asymptotic expansion of the determinant of the Hilb map, appearing in the study of critical metrics on complex manifolds via projective embeddings. This analogy helps to understand the geometric meaning of the expansion of matrix model free energy and its relation to gravitational effective actions in two dimensions. We compute the leading terms of the free energy expansion in the pure bulk case, and make some observations on the structure of the expansion to all orders. As an application of these results, we propose an asymptotic formula for the Liouville action, restricted to the space of the Bergman metrics.

\end{abstract}
\tableofcontents
\thispagestyle{empty}

\section{Introduction}

Random normal matrices are genuinely important in physics, especially in the light of their connection to the Quantum Hall effect \cite{CY, CZ, ABWZ, ZW, Zab}. They are also extremely interesting from the mathematical perspective, since they exhibit a rich geometric structure. In particular, they are related to the problem of laplacian growth and viscous flows \cite{ABWZ,TB,HM,Zab}, conformal mappings in two dimensions and integrable hierarchies \cite{W,MWZ,EF} and the Bergman kernel theory \cite{B,B1,HH}.

In this paper we further explore the connection between random normal matrices and complex geometry. The eigenvalues of normal matrices are complex-valued, which entails the relevance of this model to the geometry of the complex plane $\mathbb C$. In the series of papers \cite{B}-\cite{B3} Berman proposed a generalization of normal matrix ensembles to the situation, where $\mathbb C$ is replaced by a compact complex manifold $M$ equipped with a line bundle $L^k$, with $k$ playing the role of the large parameter $N$. The "matrix" interpretation is then lost, nevertheless the partition function has basically a similar recognizable form. The statistical sum in his model at large $k$ freezes on K\"ahler-Einstein metrics, which can be interpreted as "emergence" of gravity from the underlying statistical mechanics ensemble \cite{B4}. This observation led to many fruitful applications in the theory of K\"ahler-Einstein metrics \cite{B5}. Remarkably, relation to quantization of the graviton has been alluded to in previous work on Quantum Hall effect in higher dimensions, on $S^4$ \cite{ZH1,ZH2} and $\mathbb{CP}^k$ \cite{KN1,KN2,KN3}. The model we study in this paper is, on the one hand, closely related to Berman's model, and on the other hand is implicit already in the earlier work by Donaldson \cite{Don2} on projective embeddings. In the case of Riemann surfaces this model is also closely related to the one, studied by Zeitouni and Zelditch, using large deviation techniques \cite{ZZ,Z1}. Here we are interested in the structure of large $k$ expansion in our model, and in particular in its connection to the work by Wiegmann and Zabrodin on the large $N$ expansion for random normal matrices \cite{WZ,ZW}. 

To explain our setup, recall that in \cite{DK} we studied the quantum mechanical problem for a particle in a magnetic field on K\"ahler manifold, and derived the Bergman kernel expansion for the density matrix at the lowest Landau level. Here we construct the collective wave function of these particles, considered now as non-interacting fermions, and study the partition function of the system. We note that in the special case when $M=\mathbb CP^1$ the partition function can be represented as an integral over $N$ copies of $\mathbb C$ with special boundary conditions at infinity for the potential, thus making an explicit link with the normal matrix model. For general K\"ahler manifolds the standard determinantal representation applied to our partition function leads to the determinant of the so-called $\Hilb$ map, introduced in \cite{Don2}, in the context of Yau-Tian-Donaldson program in K\"ahler geometry \cite{Don1}. The two leading terms in the large $k$ expansion of $\det\Hilb$ were already determined in \cite{Don2}, using the Bergman kernel expansion. In this paper we pursue this analysis to higher orders in $1/k$ and make some intriguing observations on the form of this expansion to all orders.

In the random normal matrix case, corresponding to $M=\mathbb CP^1$, the state-of-the-art calculation of large $N$ free energy expansion is due to Zabrodin and Wiegmann \cite{ZW}. They used the method of Ward identities to find the first three terms in the expansion  in a more general case of beta-ensemble. This is the case of interacting fermions - random normal matrices correspond to $\beta=1$. Here we reproduce their expansion rigorously for $\beta=1$. Our result automatically holds for any Riemann surface, in the situation where there is no boundary. The K\"ahler parametrization turns out to be particularly convenient if one's goal is to understand the geometic meaning of various terms in this expansion. The first three terms correspond to the Aubin-Yau, Mabuchi and Liouville functionals. All three functionals play an important r\^ole in the problem of K\"ahler-Einstein metrics. Moreover, these functionals, most famously the Liouville action, appear as gravitational effective actions in two-dimensional quantum gravity coupled to matter \cite{FKZ1,FKZ3}, thus providing a link to the random K\"ahler metric program \cite{FKZ2}. Furthermore, we argue that the remainder term in the expansion of the free energy contains only action functionals, which are exact one-cocycles on the space of metrics, i.e. can be expressed as a difference of integrals of local density, depending on one metric. We support this conjecture by an explicit calculation up to the fifth order, using the closed formula for the Bergman kernel expansion due to Xu \cite{Xu}. We stress that our method applies only to the pure bulk situation, i.e. when the Quantum Hall droplet is supported everywhere on the manifold. One can see this as a necessary payoff for being able to look deeper into the structure of the expansion.

The paper is organized as follows. We explain our physics setup and its relation to the determinant of $\Hilb$ map in section 2. In section 3 we reduce our model to normal matrix ensemble in the special case of sphere. In section 4 we explain the method to derive the large $k$ expansion of the free energy. In section 5 we derive the first four terms in the expansion, explain their geometric meaning, relation to the Wiegmann-Zabrodin expansion, and make a conjecture about the status of the remainder terms. In section 6 we check the conjecture for the order five term. As an application, we use the results of this paper in order to construct a "quantized" (in the geometric quantization sense) version of the Liouville action, restricted to the space of Bergman metrics.

\section{Non-interacting fermions on a K\"ahler manifold}

We consider a system of non-interacting fermions in a magnetic field on a compact K\"ahler manifold $M$ of complex dimension $n$. Our setup here follows \cite{DK}. The manifold is characterized by the metric tensor $g_{a\bar b}$ with the corresponding K\"ahler form $\omega_\phi=ig_{a\bar b}dz^a\wedge d\bz^b$, being a positive-definite $(1,1)$-form. We choose the particular magnetic field configuration with the field strength proportional to the K\"ahler form 
\begin{equation}
\label{magn}
F=k\omega_\phi,
\end{equation}
where $k$ is a big integer. While more general choices of the magnetic field configuration can be considered in this context \cite{Kl}, this particular choice leads to a natural generalization of the two-dimensional lowest Landau levels to higher dimensions \cite{DK}. Mathematically, this setup corresponds to a choice of positive line bundle $L$, its tensor power $L^k$ and a hermitian metric $h^k$, The curvature $R_h=-i\p\bp\log h^k$ of the hermitian metric corresponds to the magnetic field strength. At the same time by \eqref{magn} it is proportional to the K\"ahler metric $k\omega_\phi=-i\p\bp\log h^k$. The wave functions of particles on the lowest Landau level in this background are constructed with the help of holomorphic sections $s_i$ of $L^k$, weighted by the metric on the line bundle
\begin{equation}
\label{psi}
\psi_i(z,\bz)= s_i(z)h^{\frac k2}(z,\bz),\quad i=1,..,N_k
\end{equation}
The normalization of wave functions will be chosen as follows. We assume that $\omega_\phi$ belongs to the cohomology class $[\omega_0]$ of some reference K\"ahler form $\omega_0$ with the magnetic potential $h_0^k$, meaning
\begin{equation}
\omega_\phi=\omega_0+i\p\bp\phi,
\end{equation}
where the real-valued function $\phi$ is called the K\"ahler potential. Equivalently, for the Hermitian metrics on the line bundle we have: $h^k=h_0^ke^{-k\phi}$. Now, we fix the normalization of the wave functions $\psi^0$, defined as in \eqref{psi} with respect to the reference metric $\omega_0, h_0^k$. Namely,
\begin{equation}
\label{norm}
\langle\psi^0_i,\psi^0_j\rangle=\frac1V\int_M\bs_is_j h_0^k\,\omega_0^n=\delta_{ij}.
\end{equation}
Here $V=\int_M\omega_0^n$ is the volume of $M$. It is easy to see that the volume is the same for all metrics in the cohomology class $[\omega_0]$. In what follows we normalize it as $V=(2\pi)^n$. Let us stress, that the particular choice of the reference metric  in the class $[\omega_0]$ is of no importance. The formulas below will depend on $\omega_0$ in a covariant way, so that one can easily replace $\omega_0$ by any other metric from $[\omega_0]$.

The collective wave function of $N_k$ noninteracting fermions is given by the Slater determinant
\begin{equation}
\Psi(z_1,\ldots,z_{N_k})=\frac1{\sqrt{N_k!}}\det \psi_i(z_j).
\end{equation}
The partition function per volume is then the multiple integral over the manifold of the squared norm of the wave function
\begin{eqnarray}
\label{det} 
\nonumber
Z_{N_k}&=&\frac1{V^{N_k}}\int_{M^{\otimes N_k}} |\Psi(z_1,\ldots,z_{N_k})|^2\prod_{i=1}^{N_k}\omega_{\phi}^n|_{z_i}\\ \nonumber&=&\frac1{N_k!V^{N_k}}\int_{M^{\otimes N_k}} |\det s_i(z_j)|^2 e^{-k\sum_{i=1}^{N_k}\bigl(\phi(z_i)-\log h_0(z_i)\bigr)} \prod_{i=1}^{N_k}\omega_{\phi}^n|_{z_i},\\
&=&\det\frac1V\int_M\bs_is_j h_0^ke^{-k\phi}\,\omega_\phi^n.
\end{eqnarray}
where we used Gram identity.
Here $i, j$ are matrix indices and $k$ is an (integer) parameter denoting the degree of the line bundle $L^k$.
The matrix of the size $N_k$ by $N_k$ in the last line was introduced in \cite{Don2} and is usually denoted as
\begin{equation}
\label{hilbk}
\Hilb_k(\phi)_{ij}=\frac1V\int_M\bs_i s_j h_0^ke^{-k\phi}\,\omega_{\phi}^n.
\end{equation}
Let us quickly recall its r\^ole in the K\"ahler geometry. 

Given the choice of the reference metric, the system \eqref{det} above is parameterized essentially by a single function, the K\"ahler potential $\phi$ 
\begin{equation}
Z_{N_k}=Z_{N_k}[\omega_0,\phi].
\end{equation}
The space of all K\"ahler potentials is in one to one correspondence with the space of all K\"ahler metrics on $M$ in the K\"ahler class $[\omega_0]$
\begin{equation}
\mathcal K_{\omega_0}= \{\phi\in C^{\infty}(M)/\mathbb R,\, \omega_\phi=\omega_0+i\p\bp\phi>0\},
\end{equation}
if we mod out by constant potentials, since they do not change the metric. Note that although the partition function depends on the constant mode $\phi=c$, the dependence is almost trivial
\begin{equation}
\label{const}
Z_{N_k}[\omega_0,c]=e^{-ckN_k}.
\end{equation}

Now, the holomorphic part of the wave functions \eqref{psi}, taken up to a overall scale, can be thought of as defining the embedding of the manifold $M$ to the projective space $\mathbb{CP}^{N_k-1}$, via $z\to s_i(z)$. The matrix $\Hilb_k$ 
can be understood as a map from the infinite-dimensional space $\kcalomega$ of K\"ahler potentials to the finite-dimensional space $\bcal_k$ of norms on the vector space of sections,
\begin{equation}
\label{hilbk1}
\Hilb_k:\; \mathcal K_{\omega_0}\rightarrow\bcal_k.
\end{equation}
The latter space can be identified with positive hermitian matrices of the size $N_k$ by $N_k$. We will later explain that at large $k$ the space $\bcal_k$ approximates $\kcalomega$ in a very strong sense.
%The latter can be described as a symmetric space $\bcal_k=GL(N_k,\mathbb C)/U(N_k)$.

Therefore we arrive at the following relation
\begin{equation}
\label{rel}
Z_{N_k}=\det\Hilb_k(\phi),
\end{equation} 
where again the choice of $\omega_0$ is assumed.
The normalization condition \eqref{norm} then means that the reference metric is mapped to the identity matrix in $\bcal_k$: $\Hilb_k(0)=I$. Our goal is to study the expansion of the free energy of the system \eqref{det}
\begin{equation}
\label{free}
\mathcal F=\log Z_{N_k}=\sum_{j=0}^{\infty}k^{n+1-j}\mathcal S_j(\omega_0, \phi)
\end{equation}
at large $k$. This particular form of the expansion will become obvious in what follows. In the next section we show that for $M=\mathbb{CP}^1$ the system above reduces to the normal matrix model.

\section{Relation to normal random matrices}

The simplest choice of the compact manifold is $M=\mathbb{CP}^1$. In this case the determinant of the Hilb map is related to the partition function of the normal matrix model. Indeed, let $\omega_0$ be the round metric on the sphere, and $h_0^k$ is the corresponding Hermitian metric on the line bundle. In terms of the projective coordinate $z$ we have
\begin{equation}
\omega_0=\frac{idz\wedge d\bz}{(1+|z|^2)^2},\quad  h_0^k=(1+|z|^2)^{-k},
\end{equation}
The area of $\mathbb{CP}^1$ in this metric equals $A=2\pi$.
The corresponding orthonormal basis of holomorphic sections \eqref{norm} can be constructed explicitly
\begin{equation}
s_i(z)=\sqrt{N_kC_k^{i-1}}z^{i-1},\quad i=1,\ldots,N_k=k+1.
\end{equation}
Plugging this back to the partition function \eqref{det} we immediately get
\begin{eqnarray}
\label{dethilb1}
\nonumber
Z_{N_k}&=&\frac1{(2\pi)^{N_k}N_k!}\int_{(\mathbb{CP}^1)^{N_k}}|\det s_i(z_j)|^2e^{-k\sum_{i=1}^{N_k}\bigl(\phi(z_i)-\log h_0(z_i)\bigr)}\prod_{i=1}^{N_k}\omega_\phi(z_i)\\
\nonumber
&=&\frac1{(2\pi)^{N_k}N_k!}\int_{(\mathbb{CP}^1)^{N_k}}|\det s_i(z_j)|^2e^{-k\sum_{i=1}^{N_k}\bigl(\phi(z_i)-\log h_0(z_i)-\frac1k\log\frac{\omega_\phi}{\omega_0}|_{z_i}\bigr)}\prod_{i=1}^{N_k}\omega_0(z_i)\\
&=&\frac1{\pi^{N_k}}\prod_{i=1}^{N_k}(C^{i}_{N_k})\int_{\mathbb C^{N_k}}|\Delta(z)|^2e^{-k\sum_{i=1}^{N_k}\bigl(\Phi(z_i)-\frac1k\log\frac{\p^2\Phi}{\p z\p\bz}(z_i)\bigr)}\prod_{i=1}^{N_k}d^2z_i.
\end{eqnarray}
Here $\Delta(z)=\prod_{i<j}(z_i-z_j)$ is the usual Vandermonde determinant. In the last line we effectively changed the integration domain to the $N_k$ copies of the complex plane. To this end, we introduced the planar K\"ahler potential
\begin{equation}
\label{planepot}
\Phi(z,\bz)=\phi(z,\bz)+\log(1+|z|^2),
\end{equation}
The corresponding K\"ahler form on $\mathbb C$ is just the Hessian 
\begin{equation}
\omega_\phi= \frac{\p^2\Phi}{\p z\p\bz}\,idz\wedge d\bz.
\end{equation}
The representation \eqref{dethilb1} allows us to make an explicit link with the partition function of the normal (or complex) matrix model for general eigenvalue potentials \cite{CZ,EF,HM,WZ}. We see that our $Z_{N_k}$ is proportional to the latter
\begin{equation}
\label{z}
Z_{N_k}[W]=c_k\int_{\mathbb C^{N_k}} |\Delta(z)|^2e^{-k\sum_{i=1}^{N_k}W(z_i)}\prod_{i=1}^{N_k}d^2z_i,
\end{equation}
where the normalization constant is
\begin{equation}
c_k= \frac1{\pi^{N_k}}\prod_{i=1}^{N_k}C^{i}_{N_k}.
\end{equation}
If we identify
\begin{equation}
\label{potential}
W(z)=\Phi(z)-\frac1k\log\frac{\p^2\Phi(z)}{\p z\p\bz},
\end{equation}
then the formula \eqref{rel} reduces to the usual determinantal representation of the matrix integral
\begin{equation}
Z_{N_k}[W]=\det\Hilb_k^{{\mathbb CP}^1}(\phi)=c_k\det\int_{\mathbb C}z^{i-1}\bz^{j-1}e^{-kW(z)}d^2z,
\end{equation}
compare e.g.\ to \cite{Zab}. In this light the relation \eqref{rel} can be understood as a higher-dimensional generalization of the determinantal formula.

A few comments are in order. In the case of the sphere there are several important differences compared to the standard case of the fermions in the complex plane. First, in the planar case the number of states on the lowest Landau level equals the total flux of the magnetic field, while on the sphere the number of states is $N_k=k+1$ and the magnetic flux $\int F/2\pi=k$ is one unit less. This is due to the extra normalizable mode in the compact case. Second, there is an order $1/k$ correction to the potential \eqref{potential}, which appears because of the choice of nontrivial metric \eqref{magn}. Usually one assumes the euclidean metric on $\mathbb C$ and the system is parameterized by the potential $W$. We will see that the parameterization by the K\"ahler potential is much more convenient if one's goal is to study $1/k$ expansion. However, the requirement that $\phi$ is a K\"ahler potential on $\mathbb{CP}^1$ translates into the special boundary conditions for the potential $W$. Namely, since $\phi$ shall be a bounded function everywhere on $M$, the planar K\"ahler potential $\Phi$ \eqref{planepot} behaves at infinity as $\Phi\sim\log|z|^2+\mathcal O(1)$. Therefore the potential $W$ has the following asymptotic behavior
\begin{equation}
\label{growth}
W(z)=\left(1+\frac2k\right)\log|z|^2+\mathcal O(1),\quad{\rm as}\;|z|\rightarrow\infty.
\end{equation}
Essentially, this growth at infinity is the slowest possible so that the partition function \eqref{z} converges. 
Usually in this context more general boundary conditions on the potential \cite{HM,AHM,B1,WZ} are considered
\begin{equation}
W(z)\geq(1+\epsilon)\log|z|^2+\mathcal O(1),\quad{\rm as}\;|z|\rightarrow\infty.
\end{equation}
With this growth conditions on the potential the fermions form droplets of the finite size on the plane, with the density matrix uniformly constant inside and zero outside of the boundary of the droplet. The growth conditions \eqref{growth}, which we assume in this paper, are too mild to hold the fermions together and they tend to spread out over the whole manifold ($\mathbb C$ or $\mathbb{CP}^1$). In other words the droplet has the size of the manifold and boundary does not appear. Thus we study only the bulk dynamics of the problem. The considerable advantage is that it turns out to be possible to say much more about the structure of the large $k$ expansion of the free energy.

\section{Bergman kernel method}

In the case of K\"ahler manifolds there exists a nice method to study the asymptotic expansion of the free energy \eqref{free}, or equivalently the logarithm of $\det\Hilb_k$, which we recall here, following \cite{Don2}. Consider the variation of the free energy with respect to $\phi$
\begin{eqnarray}
\nonumber
\label{dethilb}
\delta\mathcal F&=&\delta \log\det\Hilb_k(\phi)={\rm Tr} \,\Hilb_k^{-1}(\phi)\,\delta \Hilb_k(\phi)\\\nonumber&=&\frac1V\int_M\sum_{i,j=1}^{N_k}{\Hilb_k(\phi)^{-1}}_{ij}\bs_js_ih^k\bigl(-k+\Delta)\delta\phi\;\omega_\phi^n\\&=&\frac1V\int_M\bigl(-k\rho_k+\Delta\rho_k\bigr)\delta\phi\,\omega_\phi^n.
\end{eqnarray}
Here ${\Hilb_k(\phi)}^{-1}$ is the inverse of the matrix $\Hilb_k(\phi)$ and the laplacian $\Delta$ is minus one-half of the ordinary riemmanian laplacian, taken in the metric $\omega_\phi$. The sum in the second line
\begin{equation}
\rho_k(z)=\sum_{i,j=1}^{N_k}{\Hilb_k(\phi)^{-1}}_{ij}\bs_js_ih^k
\end{equation}
is known as the Bergman kernel, in this case restricted to the diagonal. This is the key object of our study. There exists \cite{T,Z,C,Lu} the following asymptotic large $k$ expansion 
\begin{equation}
\label{TYZ}
\rho_k(z)=
 k^n+\frac12k^{n-1}R+k^{n-2}\left(\frac13\Delta
 R+\frac1{24}|\mbox{Riem}|^2-\frac16|\mbox{Ric}|^2+\frac18R^2\right)
 +\CO(k^{n-3}),
\end{equation}
see also \cite{MM,MM1} for a review.
The expansion on the rhs depends only on the Riemann, Ricci tensors and the scalar curvature and their covariant derivatives, all these quantities taken in the metric $\omega_\phi$. We adopt here standard conventions in K\"ahler geometry
\begin{eqnarray}
\nonumber
{\rm Riem}_{i\bar j l\bar m}=\p_l\bp_{\bar m}g_{i\bar j}-g^{p\bar q}\p_lg_{i\bar q}\bp_{\bar m}g_{p\bar j}\\
R_{i\bar j}=-\p_i\bp_{\bar j}\log\det g,\quad R=g^{i\bar j}R_{i\bar j}.
\end{eqnarray}
In this conventions $R$ is one half the scalar curvature in riemannian geometry.

By construction, the integral over the Bergman kernel over $M$ gives the total number of sections, which is a polynomial in $k$ of degree $n$ given by the Riemann-Roch formula 
\begin{eqnarray}
\nonumber
\label{number}
N_k=\frac1V\int_M\rho_k(z)\omega_\phi^n&=&\frac1V\int_M{\rm ch}(L^k){\rm Td}(M)\\&=&k^n+\frac12k^{n-1}c_1(M)+...=k^n+\frac12k^{n-1} \bR+...
\end{eqnarray}
We introduced the average curvature of the manifold as $\bR=\frac1V\int_MR\,\omega_\phi^n$. It is an invariant for all metrics in $\kcalomega$, and thus is a $\phi$-independent constant.

Now we can plug the Bergman kernel expansion to Eq. \eqref{dethilb} and integrate the variational formula order by order in $1/k$, taking into account the base-point condition 
\begin{equation}
\label{basept}
\det\Hilb_k(0)=1.
\end{equation}
This method essentially is a covariant generalization of the Bergman kernel method in the normal matrix model. Recall that in the latter case it is customary to work with the density $\rho(z)=\frac1N\sum\delta(z-z_i)$. Its average equals the variation of the free energy with respect to the potential $W$
\begin{equation}
\langle\rho(z)\rangle=N^2\frac{\log Z_N}{\delta W(z)}.
\end{equation}
Our definition of the kernel $\rho_k(z)$ differs from $\langle\rho(z)\rangle$, but the analogy with \eqref{dethilb} is clear. The advantage of the K\"ahler potential parameterization of the free energy is in the availability of an independent method to extract the large $k$ expansion (\ref{TYZ}) of $\rho_k(z)$. In the next section we apply the Bergman kernel method in order to determine the structure of large $k$ expansion of the free energy in complex dimension one, i.e. on any compact Riemann surface.

\section{Large $k$ expansion in complex dimension one}

Let us return to complex dimension one, which is where the original problem of non-interacting fermions takes place. In this case we can go a few orders higher in the $1/k$ expansion. For $n=1$ the Bergman kernel expansion involves only the scalar curvature and its derivatives. We list all the terms in the expansion up to the fifth term
\begin{eqnarray}
\label{bk3}
\rho_k(z)&=& k+\frac12R+\frac1{3k}\Delta
 R+\frac1{k^2}\left(\frac18\Delta^2 R-\frac5{48}\Delta(R^2)\right)\\\nonumber
&&+\frac1{k^3}\bigl(a_1\Delta(R^3)+a_2\Delta^2(R^2)+a_3\Delta^3R+a_4\Delta(R\Delta R)\bigr)+\CO(1/k^4).
\end{eqnarray}
The first three terms here can be immediately read off from \eqref{TYZ}, taking into account that in $n=1$ one has $|\mbox{Ric}|^2=|\mbox{Riem}|^2=R^2$. The fourth term (order $k^{n-3}$ in the expansion \eqref{TYZ}) was computed for any $n$ in \cite{Lu}. Here we present it in the form, specified to $n=1$. Note that the terms with negative powers of $k$ must be full derivatives since the Riemann-Roch formula \eqref{number} counting the number of sections terminates at the order $k^0$. This explains the structure of the fifth term in the expansion above. At this order there exists only four independent metric invariants. We choose here a convenient basis for these invariants, with corresponding numerical coefficients $a_i$ to be determined later. Also for simplicity we normalize the area of the metrics in $\kcalomega$ as $A=\int_M\omega_0=2\pi$, and explain the convertion back to arbitrary $A$ in the Appendix.

Plugging the expansion \eqref{bk3} in the formula \eqref{dethilb} we then integrate out the variational formula taking into account the base-point condition \eqref{basept}. We then get the first four terms in the asymptotic expansion of the free energy
\begin{eqnarray}
\label{exp1}
\nonumber
\mathcal F&=&\log\det\Hilb_k(\phi)\\&=&\nonumber
-2\pi kN_kS_{AY}(\omega_0,\phi)+\frac k2S_M(\omega_0,\phi)+\frac1{12\pi}S_L(\omega_0,\phi)\\
&&-\frac5{96\pi k}\left(\int_MR^2\,\omega_\phi-\int_MR_0^2\,\omega_0\right)+\mathcal O(1/k^2).
\end{eqnarray}
Let us now explain the ingredients that enter on the right hand side.
The number of sections $N_k$ \eqref{number} for a Riemann surface of genus $h$ is simply 
\begin{equation}
N_k=k+1-h.
\end{equation}
It follows that the average curvature is $\bR=2(1-h)$. Note that we rearranged the order $k^2$ and order $k$ terms to get the $N_k$ structure in front of the first term.
The first two functionals here are known in K\"ahler geometry as the Aubin-Yau and Mabuchi actions, correspondingly. In dimension one they read 
\begin{eqnarray}
S_{AY}(\omega_0,\phi)&=&\frac1{(2\pi)^2}\int_M\left(\frac12\phi\Delta_0\phi+\phi\right)\omega_0,\\
S_M(\omega_0,\phi)&=&\frac1{2\pi}\int_M\frac{\bR}2\phi\Delta_0\phi\,\omega_0+\phi\bigl(\bR \omega_0-\mbox{Ric}(\omega_0)\bigr)+\omega_\phi\log\frac{\omega_\phi}{\omega_0},
\end{eqnarray}
where $\mbox{Ric}(\omega_0)=\mbox{Ric}_{z\bz}(\omega_0) idz\wedge d\bz =R_0\omega_0$ is the Ricci form in the metric $\omega_0$ and $R_0$ is the scalar curvature of $\omega_0$. Note that the leading term for constant K\"ahler potentials equals
\begin{equation}
\label{const1}
S_{AY}(\omega_0,c)=c/2\pi,
\end{equation} 
in complete agreement with \eqref{const}.

While the variation of the Aubin-Yau functional is just the volume form, the variation of the Mabuchi functional gives the constant scalar curvature condition
\begin{equation}
\label{mabcrit}
\delta S_M(\omega_0,\phi)=\frac1{2\pi}\int_M(\bR-R)\,\delta\phi\,\omega_\phi.
\end{equation}

The third functional in \eqref{exp1} is the Liouville action
\begin{equation}
\label{L}
S_L(\omega_0,\phi)=\int_M\frac12\bigl(\mbox{Ric}(\omega_\phi)+\mbox{Ric}(\omega_0)\bigr)\log\frac{\omega_\phi}{\omega_0}.
\end{equation}
This functional is more familiar in physics when written in terms of the conformal Weyl field $\sigma$, which can be identified with the K\"ahler potental as $e^{2\sigma}=1+\Delta_0\phi$. In this parametrization of the metric we have
\begin{equation}
\label{liouv1}
S_L(\omega_0,\sigma)=\int_M\bigl(-2\sigma\p\bp\sigma+2\mbox{Ric}(\omega_0)\sigma\bigr)=\int_M\bigl(g_0^{ij}\p_i\sigma\p_j\sigma+2R_0\sigma\bigr)\sqrt{g_0}d^2x
\end{equation}
Note that the identification between the conformal and K\"ahler fields above implies the constant area constraint for the cosmological term $\int_Me^{2\sigma}\omega_0=A=2\pi$, since all K\"ahler metrics have the same area. Therefore \eqref{liouv1} is written by default in the fixed area gauge. The variation of the Liouville action with respect to K\"ahler potential 
\begin{equation}
\delta S_L(\omega_0,\phi)=\int_{M}(\Delta R)\,\delta\phi\,\omega_\phi
\end{equation}
in the compact case reproduces the constant scalar curvature equation on the metric $\omega_\phi$ 
\begin{equation}
R=\bR,
\end{equation}
which is the same equation of motion as for the Mabuchi action.

It is instructive to think of the functionals above as depending on a pair of metrics $\omega_0,\omega_\phi$ in $\kcalomega$. The only caveat here is that one has to keep in mind that the Aubin-Yau action depends on the constant mode of $\phi$ in a simple way \eqref{const1}. Tautologically, we have
\begin{equation}
S(\omega_0,\phi)\equiv S(\omega_0,\omega_\phi).
\end{equation}
Then an interesting property of the expansion can be immediately pointed out. All the functionals entering the expansion \eqref{exp1} satisfy the so-called cocycle identity. Namely, for any triplet of metrics $\omega_0,\omega_1,\omega_2\in\kcalomega$ we have
\begin{eqnarray}
\label{cocyc}
S(\omega_0,\omega_2)=S(\omega_0,\omega_1)+S(\omega_1,\omega_2),
\end{eqnarray}
and also the antisymmetry $S(\omega_0,\omega_1)=-S(\omega_1,\omega_0)$. One can verify this property for the Aubin-Yau, Mabuchi and Liouville actions independently by an explicit, if tedious, calculation. Formally it follows immediately from the fact that the variation \eqref{dethilb} depends only on the metric $\omega_\phi$ and not on $\omega_0$, and from the base-point condition \eqref{basept}, which imposes antisymmetry. For the fourth term the Eq. \eqref{cocyc} holds trivially, since it is an explicit difference $S(\omega_0,\omega_1)=S(\omega_1)-S(\omega_0)$. 
It follows that the cocycle property holds for the free energy as  whole, meaning
\begin{equation}
\mathcal F(\omega_0,\omega_2)= \mathcal F(\omega_0,\omega_1)+ \mathcal F(\omega_0,\omega_2). 
\end{equation}
In particular, one can use this relation in order to replace the reference metric $\omega_0$ in the definition of the free energy by another reference metric $\omega_1$ in a controlled fashion. 
In the context of 2d gravity this property of gravitational effective actions is important ingredient behind the background independence \cite{FKZ3}.

Let us now comment on the relation between the expansion \eqref{exp1} and the work \cite{ZW}, where first three terms in the large $N$ expansion of the free energy of the Dyson gas were derived. In order to compare the answers we should express our result in terms of $W$ by solving the relation \eqref{potential} for $\phi$ order by order in $1/k$
\begin{equation}
\phi=W+\log h_0+\frac1k(\log\Delta_0W+2\log h_0)+\frac1{k^2}\frac1{\Delta_0W}\bigl(\Delta_0\log\Delta_0W-2\bigr)+...
\end{equation}
and plug it back to the expansion \eqref{exp1}. We get
\begin{eqnarray}
\mathcal F&=&-\frac{k^2}{2\pi}\int W\Delta_0 W\omega_0+\frac k{4\pi}\int\bigl(\bar RW -\Delta_0W\log\Delta_0W\bigr)\omega_0+\frac1{12\pi}S_L(\omega_0, \Delta_0W)+...
\end{eqnarray}
The first two terms here coincide with the corresponding terms in \cite{ZW}, for $\beta=1$ and assuming no boundary. To compare the last term, recall that the Liouville action is related to the determinant of laplacian (conformal anomaly) as follows
\begin{equation}
\label{detlaplacian}
\log\frac{\det\Delta_{e^{2\sigma}\omega_0}}{\det\Delta_0}=-\frac1{12\pi}S_L(\omega_0,\sigma).
\end{equation}
Therefore this term corresponds to the logarithm of the determinant of the laplacian in \cite{ZW}. Thus our method here provides a first rigorous derivation of this term. 

Finally, a few comments on the relation of the above expansion to two-dimensional gravity, alluded to already in \cite{ZW}. The free energy $\mathcal F$ can be thought of as a generating function for the geometric functionals, appearing as gravitational effective actions in 2d gravity. The Liouville action appears in the Polyakov 2d quantum gravity coupled to conformal matter. The Mabuchi functional, implicit already in \cite{ZW}, appears when the matter theory is disturbed by the mass term \cite{FKZ3} (the Aubin-Yau action is just a part of the Mabuchi action). The quadratic curvature term has also been considered in quantum 2d gravity in \cite{Kaz}.

\section{The remainder term as an exact one-cocycle}

Another interesting feature of the expansion \eqref{exp1} is that the order $1/k$ term turns out to be a exact one-cocycle in $\kcalomega$. We say that the one-cocyle is exact, or equivalently, is a coboundary, if it can be written as a difference
\begin{equation}
\label{diff}
S(\omega_0,\omega_\phi)=\tilde S(\omega_\phi)-\tilde S(\omega_0),
\end{equation}
for a {\it local} functional $\tilde S(\omega)$ of the metric, i.e.\ a single integral of a local  density, depending only on curvature invariants of the metric $\omega$, and on the covariant derivatives thereof. For instance, the Liouville action on $\mathbb C$ can be written as a difference \eqref{diff}, but only with the {\it nonlocal} Polyakov action $S=\iint R\frac1\Delta R$. Therefore it is not an exact cocycle.

This special structure of the order $1/k$ term in the expansion \eqref{exp1} can be traced back to a particular combination of $1/k$ and $1/k^2$ order terms in the expansion of the Bergman kernel \eqref{TYZ}, which together produces a simple $R^2$ structure in the free energy expansion. It is natural to conjecture, that all of the remainder terms, i.e. terms in the expansion \eqref{free} starting from order $1/k$, are exact one-cocycles. The terms with $j>2$ in the expansion \eqref{free} (in dimension $n=1$) should then be the differences of the type \eqref{diff} of integrals over various curvature invariants
\begin{equation}
\tilde S_j(\omega_\phi)=\int_M \bigl(b_1R^{j+1}+b_2R^{j-1}\Delta R+...\bigr)\omega_\phi
\end{equation}
This condition leads to some new constraints on the coefficients of the Bergman kernel expansion \eqref{bk3}, mixing different coefficients in its expansion. For instance, the most general local functional, which can appear at the order $1/k^2$  in the free energy expansion, is the sum of only two terms
\begin{equation}
\tilde S_2=\int_M\bigl(b_1R^3+b_2R\Delta R\bigr)\omega_\phi,
\end{equation}
with two unknown coefficients $b_1,\,b_2$, whereas the Bergman kernel expansion at this order depends on four coefficients $a_1,a_2,a_3$ and $a_4$ \eqref{bk3}. Therefore if true, our conjecture leads to the following {\it apriori} constraints on these coefficients
\begin{equation}
\label{apri}
\frac32a_1=a_2+\frac5{48},\quad
a_3=a_4+\frac18.
\end{equation}
In order to check our conjecture for the next order term we computed the fourth-order coefficients of the Bergman kernel expansion in complex dimension one using the graph-theoretic formula \cite{Xu}. One has to sum up the contributions from the corresponding Feynman graphs listed in the appendix of \cite{Xu}. The calculation greatly simplifies for $n=1$ due to small number of curvature invariants. Here we present the final result, writing down the Bergman kernel expansion to this order in complex dimension one
\begin{eqnarray}
\label{bk4}
\rho_k(z)&=& k+\frac12R+\frac1{3k}\Delta
 R+\frac1{k^2}\left(\frac18\Delta^2 R-\frac5{48}\Delta(R^2)\right)\\\nonumber
&&+\frac1{k^3}\left(\frac{29}{720}\Delta(R^3)-\frac{7}{160}\Delta^2(R^2)+\frac{1}{30}\Delta^3R-\frac{11}{120}\Delta(R\Delta R)\right)+\CO(1/k^4).
\end{eqnarray}
One can immediately check that the last term is in a perfect agreement with the constraints \eqref{apri}. The free energy expansion to this order reads
\begin{eqnarray}
\label{exp2}
\nonumber
\mathcal F&=&\log\det\Hilb_k(\phi)\\&=&\nonumber-2\pi kN_kS_{AY}(\omega_0,\phi)+\frac k2S_M(\omega_0,\phi)+\frac1{12\pi}S_L(\omega_0,\phi)\\
&&-\frac5{96\pi k}\left(\int_MR^2\,\omega_\phi-\int_MR_0^2\,\omega_0\right)\\\nonumber&&
+\frac1{2880\pi k^2}\left(\int_M(29R^3-66R\Delta R)\,\omega_\phi-\int_M(29R_0^3-66R_0\Delta_0 R_0)\, \omega_0\right)+\mathcal O(1/k^3),
\end{eqnarray}
in perfect agreement with our conjecture.

Let us now comment on the situation in complex dimension $n$. The natural extension of the conjecture to dimension $n$ is that the terms with negative powers of $k$, i.e. terms with $j>n+1$ in the expansion \eqref{free} shall be exact one-cocycles. The expansion starts at the order $k^{n+1}$. From this we deduce that the first $n+2$ terms must be some nontrivial action functionals. It would be interesting to identify the set of relevant functionals, e.g.\ in terms of the known Chen-Tian energy functionals on $\kcalomega$, see \cite{CT}. Let us compute the expansion up to the order $k^{n-1}$. We get
\begin{eqnarray} 
\label{Fn}
\nonumber
\delta\mathcal F&=&\delta\log\det\Hilb_k(\phi)=-2\pi kN_k\delta S_{AY}(\omega_0,\phi)+\frac{k^n}2\delta S_M(\omega_0,\phi)\\&+&
k^{n-1} \int_M \left(\frac16\Delta
 R-\frac18R^2+\frac16|\mbox{Ric}|^2-\frac1{24}|\mbox{Riem}|^2\right)\,\delta\phi\,\omega_\phi^n+\mathcal O(k^{n-2})
\end{eqnarray}
The first two terms here were computed already in \cite{Don2}. They are the Aubin-Yau and Mabuchi functionals, defined in dimension $n$ by the following variational formulas
\begin{eqnarray}
\label{AY}
\delta S_{AY}(\omega_0,\phi)&=&\frac1{2\pi V}\int_M\delta\phi\,\omega_\phi^n\\
\delta S_M(\omega_0,\phi)&=&\frac1V\int_M(\bR-R)\delta\phi\,\omega_\phi^n.
\end{eqnarray}
For explicit formulas in complex dimension $n$ we refer the reader to \cite{PS}. 
The order $k^{n-1}$ term in \eqref{Fn} involves four different curvature structures. The combination of $\Delta R$, $R^2$ and $|\mbox{Ric}|^2$ enters the $E_1$ and $E_2$ functionals in \cite{CT}. However, the last term is a new functional in K\"ahler geometry
\begin{equation}
\delta S_{\small{\rm Riem}}=\frac1V\int_M|\mbox{Riem}|^2\,\delta\phi\,\omega_\phi^n
\end{equation}
Therefore we conclude that the known set of energy functionals may be insufficient for the reconstruction of the free energy expansion in complex dimension $n$.

\section{Liouville action restricted to the Bergman metrics}

Another application of the expansion \eqref{exp2} is its relation to the random K\"ahler metrics program \cite{FKZ1}-\cite{FKZ4}. 
Recall that the matrix $\Hilb_k(\phi)$ can be thought of as a map \eqref{hilbk} from the space of K\"ahler potentials $\kcalomega$ to the positive definite hermitian matrices $\bcal_k$. Given a matrix $P$ from $\bcal_k$, there exists a map in the opposite direction $FS_k(P):\bcal_k\rightarrow\kcalomega$, which is constructed as follows
\begin{equation}
\phi(P)=FS_k(P)=\frac1k\log\bar s_i{P^{-1}}_{ij}s_jh_0^k,
\end{equation}
and the corresponding K\"ahler metric 
\begin{equation}
\label{bergm}
\omega_{\phi(P)}=\omega_0+i\p\bp FS_k(P)=\frac1ki\p\bp\log\bar s_i{P^{-1}}_{ij}s_j
\end{equation}
is called the Bergman metric. These two maps are in general not mutually inverse. However, the composition of the two maps 
\begin{equation}
\label{composition}
FS_k\circ \Hilb_k(\phi)=\phi+\frac1k\log\rho_k
\end{equation}
is very close to identity in the sense that
\begin{equation}
\omega_{\phi(\Hilb_k(\phi))}-\omega_\phi=\mathcal O(1/k^2)
\end{equation}
as follows from the expansion \eqref{TYZ} of the Bergman kernel. Therefore for any K\"ahler potential $\phi$ and the corresponding K\"ahler metric $\omega_\phi$, there exists a Bergman metric asymptotically close to $\omega_\phi$ as $k\rightarrow \infty$. This is the metric \eqref{bergm} with the matrix $P=\Hilb_k(\phi)$. In other words, any metric from $\kcalomega$ can be approximated by Bergman metrics with arbitrary precision at large $k$. This statement, known as the Tian-Yau-Zelditch theorem \cite{T,Z}, essentially means that $\kcalomega=\lim_{k\rightarrow\infty}\bcal_k$.

In one of the approaches to random K\"ahler metrics (the bottom-up approach), formulated in \cite{FKZ2} we suggest to treat the positive hermitian matrix 
$$P=\Hilb_k(\phi)$$ as a random (matrix) variable. One particularly nice feature of this approach is the fact that the natural Mabuchi-Semmes-Donaldson metric on $\kcalomega$ turns out to correspond simply to the pull-back of the simplest possible invariant metric on $\bcal_k$. The hard part is then to construct approximations to geometric action functionals such as Mabuchi and Liouville actions. Donaldson  \cite{Don2} used the formula \eqref{Fn} to build the approximaion to the Mabuchi functional on the space of Bergman metrics. Indeed, from \eqref{composition} it follows immediately, that the functional
\begin{equation}
\label{bal1}
S_k(P)=2\pi kN_kS_{AY}(\omega_0,\phi(P))-\log\det P=\frac{k^n}2S_M(\omega_0,\phi(P))+\mathcal O(k^{n-1})
\end{equation}
approximates the Mabuchi energy, restricted to the space of Bergman metrics. The functional 
$S_k$ is called the balancing energy. Recall, that the critical point of the Mabuchi functional corresponds to the constant scalar curvature metric \eqref{mabcrit}. The critical point of the balancing energy is the balanced metric, which, when it exists, approximates the constant scalar curvature metric \cite{Don1} at large $k$. The convergence of the Hessian of the balancing energy was established in \cite{F}.

Having established the formula \eqref{exp2}, we can now construct the balancing energy for the Liouville functional by analogy with the balancing energy \eqref{bal1} for the Mabuchi functional. In complex dimension one we have
\begin{equation}
S_{L,k}(P)=2\pi kN_kS_{AY}(\omega_0,\phi(P))-\log\det P-\frac k2S_M(\omega_0,\phi(P))=\frac1{12\pi}S_L(\omega_0,\phi(P))+\mathcal O(1/k)
\end{equation}
Essentially the Liouville balancing energy is the difference between the balancing energy and the Mabuchi action, restricted to the Bergman metrics.
\begin{equation}
\label{bal2}
S_{L,k}(P)=S_k(P)-\frac k2S_M(\omega_0,\phi(P)).
\end{equation}
 Thus we managed to construct the approximation on $\bcal_k$ to the Liouville action in terms of more simple functionals, whose behavior on the space $\bcal_k$ is understood much better \cite{PS}.

\section{Discussion}

The main technical result of this paper is the rigorous derivation of expansion \eqref{exp2} of the free energy of non-interacting fermions on compact Riemann surfaces with K\"ahler potential $\phi$. For the sphere the first three terms agree with the previous result by Zabrodin and Wiegmann \cite{ZW} for beta-ensembles (at $\beta=1$). For general K\"ahler manifolds this expansion is related to the determinant of the Donaldson's $\Hilb$ map, which plays an important r\^ole in K\"ahler geometry \cite{Don2}. It would be extremely interesting to generalize our results to general beta-ensembles, in order to understand their interplay with geometry, see \cite{B4} for the first steps in this direction.

We also argued that the remainder term in the free energy expansion in dimension $n=1$ contains only exact one-cocycle functionals, and checked this by an explicit calculation up to the order $1/k^2$. This observation imposes new constraints on the coefficients of the Bergman kernel expansion. Let us also point out, that using the relation \eqref{dethilb} between the free energy and the Bergman kernel, we have at least in principle a hold on the free energy expansion to all orders, given the Xu's closed formula \cite{Xu} for all-order Bergman kernel expansion via Feynman diagrams. The situation here is somewhat similar to the Hermitian 1-matrix model, where the free energy can be determined to all orders also by diagrammatic techniques \cite{CE}.

On the conceptual level, we hope our results shed a new light on the
relation between large $N$ matrix models and complex geometry.

%\bigskip

{\bf Acknowledgments}. The author is grateful to A. Alexandrov, R. Berman, M. Douglas, F. Ferrari, G. Marinescu, N. Orantin, P. Wiegmann and S. Zelditch for useful discussions. He would also like to thank the International Institute of Physics, Natal, Brazil, where this paper was completed, for providing the visiting fellowship and for hospitality. 

The author is supported by the postdoctoral fellowship from the Alexander von Humboldt Foundation. He is also supported in part by the DFG-grant ZI 513/2-1, grants RFBR 12-01-00482, RFBR 12-01-33071 (mol$\_$a$\_$ved), NSh-3349.2012.2, and by Ministry of Education and Science of the Russian Federation under the contract 8207.

\section*{Appendix: Dependence on the area}

In the main text we work for simplicity with the fixed area $A=2\pi$. To facilitate the dimensional analysis and for future applications it is instructive to restore the dependence on $A$ in the main formulas. Here we assume that the metrics in $\kcalomega$ have an arbitrary area $A$. Then
\begin{equation}
\omega_\phi=\omega_0+Ai\p\bp\phi,
\end{equation}
and $\phi$ is now a dimensionless variable. The Bergman kernel expansion reads
\begin{eqnarray}
\rho_k(z)&=& k+\frac{A}{4\pi k}R+\frac1{3k}\left(\frac{A}{2\pi}\right)^2\Delta
 R+\frac1{k^2} \left(\frac{A}{2\pi}\right)^3\left(\frac18\Delta^2 R-\frac5{48}\Delta(R^2)\right)\\\nonumber
&&+\frac1{k^3} \left(\frac{A}{2\pi}\right)^4\left(\frac{29}{720}\Delta(R^3)-\frac{7}{160}\Delta^2(R^2)+\frac{1}{30}\Delta^3R-\frac{11}{120}\Delta(R\Delta R)\right)+\CO(1/k^4),
\end{eqnarray}

The free energy in complex dimension one reads
\begin{eqnarray}
\nonumber
\mathcal F&=&\log\det\frac1A\int_M\bs_is_j h_0^ke^{-2\pi k\phi}\,\omega_\phi=
-2\pi kN_kS_{AY}(\omega_0,\phi)+\frac k2S_M(\omega_0,\phi)+\frac1{12\pi}S_L(\omega_0,\phi)\\
&&-\frac5{96\pi k}\frac A{2\pi}\int_M\bigl(R^2\,\omega_\phi-R_0^2\,\omega_0\bigr)\\\nonumber&&
+\frac1{2880\pi k^2}\left(\frac{A}{2\pi}\right)^2\int_M\left((29R^3-66R\Delta R)\,\omega_\phi-(29R_0^3-66R_0\Delta_0 R_0)\, \omega_0\right)+\mathcal O(1/k^3),
\end{eqnarray}
where the Aubin-Yau and Mabuchi action functionals are now given by
\begin{eqnarray}
\nonumber
S_{AY}(\omega_0,\phi)&=&\int_M\left(\frac12\phi\Delta_0\phi+\frac\phi A\right)\omega_0,\\
S_M(\omega_0,\phi)&=&\int_M\frac12 A\bR\phi\Delta_0\phi\,\omega_0+(\bR\omega_0-\mbox{Ric}(\omega_0))\phi+\frac{\omega_\phi}{A}\log\frac{\omega_\phi}{\omega_0}
\end{eqnarray}
and now $A\bR=4\pi(1-h)$, and the Liouville action has exactly the same area-independent form \eqref{L}. 
In order the restore the fixed-area results, one should put $A=2\pi$ and rescale $\phi\rightarrow\phi/2\pi$ in the formulas above.

%\vspace{1 cm}

\end{document}